# PASS: A Parameter Audit-based Secure and Fair Federated Learning Scheme against Free-Rider Attack

Jianhua Wang, Xiaolin Chang, *Senior Member*, *IEEE*, Jelena Mišić, *Fellow*, *IEEE*, Vojislav B. Mišić, *Senior Member*, *IEEE*, and Yixiang Wang

*Abstract*—Federated Learning (FL) as a secure distributed learning framework gains interests in Internet of Things (IoT) due to its capability of protecting the privacy of participant data. However, traditional FL systems are vulnerable to Free-Rider (FR) attacks, which causes unfairness, privacy leakage and inferior performance to FL systems. The prior defense mechanisms against FR attacks assumed that malicious clients (namely, adversaries) declare less than 50% of the total amount of clients. Moreover, they aimed for Anonymous FR (AFR) attacks and lost effectiveness in resisting Selfish FR (SFR) attacks.

In this paper, we propose a Parameter Audit-based Secure and fair federated learning Scheme (PASS) against FR attack. PASS has the following key features: (a) prevent from privacy leakage with less accuracy loss; (b) be effective in countering both AFR and SFR attacks; (c) work well no matter whether AFR and SFR adversaries occupy the majority of clients or not. Extensive experimental results validate that PASS: (a) has the same level as the State-Of-The-Art method in mean square error against privacy leakage; (b) defends against AFR and SFR attacks in terms of a higher defense success rate, lower false positive rate, and higher F1-score; (c) is still effective where adversaries exceed 50%, with F1-score 89% against AFR attack and F1-score 87% against SFR attack. Note that PASS produces no negative effect on FL accuracy when there is no FR adversary.

## I. INTRODUCTION

With the growing attention to data privacy, Federated Learning (FL), a powerful and secure distributed machine learning paradigm, is being widely used in Internet of Things (IoT) [1]–[4]. It is formulated as a multi-round model-training strategy between multiple agents. FL participants, such as remote laptops and edge IoT devices controlled by a central server, co-maintain a global model without sharing their private datasets. The standard procedure of conventional FL is illustrated in the left part of **Fig. 1**. Concretely, in **Step 1**, the central server allocates the initialized model to each participant. In **Step 2**, FL participants train the global model via their private dataset. In **Step 3**, FL participants upload the local update to the central server. In **Step 4,** the server adopts the model aggregation algorithms to aggregate the local update. In **Step 5**, FL participants receive a new global model update from the central server to continue the subsequent training process until the jointly trained model converges.

J. Wang, X. Chang and Y. Wang are with Beijing Key Laboratory of Security and Privacy in Intelligent Transportation, Beijing Jiaotong University, China. E-mail: {20112051, xlchang, 18112047}@bjtu.edu.cn.

J. Mišić and V. B. Mišić are with Ryerson University, Toronto, ON, Canada M5B 2K3. E-mail: {jmisic, vmisic} @ryerson.ca.

The past years witnessed various attacks reported in conventional FL systems and model aggregation algorithms, such as Free-Rider (FR) attacks [5]–[7]. This type of attack allows FR adversaries to benefit from a well-trained global model without contributing their own private datasets and computing resources. The right part of **Fig. 1** illustrates an FL scenario under FR attacks, where an FR adversary uploads a fake local update to the central server in **red step 3**. This upload can lead to the emergence of opportunistic behaviors and then the following severe threats: (a) the jointly trained model is more susceptible to reducing the accuracy of specific tasks [8]; (b) FR adversaries obtain the leakage gradients or local updates to reconstruct the personal data [9], [10] and then conduct model inversion [11]; (c) some FR attacks can lead to unfair training and then lower the devotion aspiration of honest participants [8].

According to whether an FR adversary dominates private data and computing power, FR attacks are categorized into two types [12]: Anonymous Free-Rider (**AFR**) and Selfish Free-Rider (**SFR**) attacks. AFR denotes that adversaries do not own any private datasets and computation resources. It is a generic form of Gaussian attack [13], where an AFR adversarial client uploads its stochastic Gaussian noise to the central server [5]. SFR means that adversaries have their own private dataset and training ability but they are unwilling to devote their data and computation resources to global model training. SFR attack gives rise to a new threat to FL.

Researchers have proposed methods to resist FR attacks, including contribution-based and robust model aggregation-based methods [8]. However, the existing works only concentrate on resisting AFR attack and suffer from the following shortcomings:

- They lose defense capability when adversaries are the majority in the system [14]. Namely, their methods usually make strong assumptions that most FL clients are honest [14], [15].
- It is hard, if not impossible, for them to resist SFR attack. Compared with AFR attack, SFR attack is more sequestered and challenging to be identified [12]. The common contribution-based methods depend on clients' contributions by calculating cosine similarity between the global update and local update [16]. However, similarities may make low contributions to high-quality workers [17], especially in SFR attack scenarios [12]. Moreover, robust model aggregation-based methods can not eliminate FR adversaries since they only focus on overall performance. Hence, they misclassify fair clients into FR adversaries and cause a higher false positive rate.

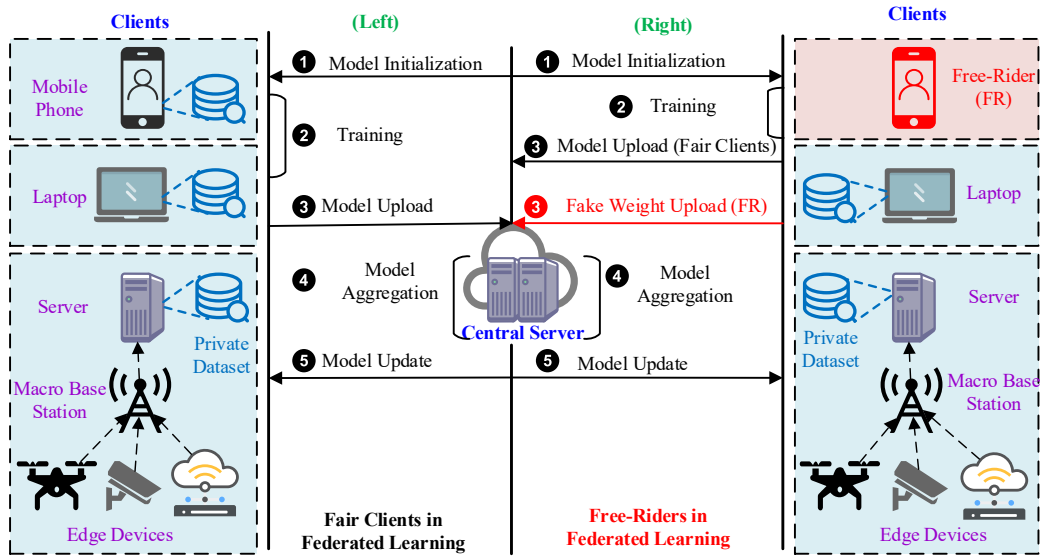

**Fig. 1.** The schematic diagram of federated learning with fair clients and free-rider clients.

Motivated by the above discussions, we propose a *P*arameter *A*udit-based *S*ecure and fair FL *S*cheme (**PASS**) against both AFR and SFR attacks in this paper, which can not only address the aforementioned problems in AFR attack-oriented defense mechanisms but also resist the SFR attack. PASS contains two components: (1) contribution evaluation (denoted as **PASS-CE**) with the aim to audit parameters, and (2) a privacy-preserving strategy (denoted as **PASS-PPS**) combining the weak Differential Privacy [18] with Gaussian mechanism and parameter prune approach.

To the best of our knowledge, we are not only the first to propose the parameter audit-based method against the AFR attack, but also the first to investigate the defense in SFR attack scenarios. We summarize the key features of PASS as follows:

- PASS can protect private data with less accuracy loss. It is achieved by the component PASS-PPS with variance $10^{-2}$ Gaussian perturbation. PASS-PPS has lower than 2% accuracy loss in Cifar10 non-independent and identically distributed (iid) and 3% in MNIST non-iid.
- PASS can defend against both AFR and SFR attacks in FL systems. It is achieved by the component PASS-CE. It focuses on individual performance intuitively and then it can effectively distinguish FR adversaries from fair clients.
- PASS works well even if AFR and SFR adversaries make up the majority of the total amount of clients. The existing defense models lose effectiveness when AFR adversaries are more than 50% in the FL system [14], [15].

Notably, frequent communications between the server and clients in the parameter audit phase will cause more time consumption. To address this problem, we adopt synchronous transmission to decrease communication times and use parameter prune to sparse local updates, and then reduce communication overhead. We analyze the overhead reduction theoretically in Section III.D.

Extensive experiments are carried out to evaluate the effectiveness of PASS and explore the proper hyper-parameters of PASS in Cifar10 and MNIST with iid and non-iid data. Notably, the commonly used model aggregation algorithm, Federated Averaging (FedAvg), is adopted. The experiment results demonstrate that:

- PASS has the same level as the State-Of-The-Art (SOTA) method [16] in mean square error against privacy leakage, as shown in **Fig.8**.
- Compared with other FR attack defense models [8], [15], [19], PASS can achieve a 100% Defense Success Rate (DSR), 20% False Positive Rate (FPR), and the highest F1-score (on average) of 88% against SFR and 89% against AFR, which are the best performance compared with other defense schemes [8], [15], [19].
- PASS achieves better defense performance no matter whether AFR and SFR adversaries occupy the majority of clients or not. Especially in the SFR attack, PASS obtains the highest F1-score of 87% in adversaries more than 50% scenario and 89% in adversaries less than 50%, illustrated in **Fig.13**.

The rest of the paper is organized as follows. Section II presents preliminary knowledge and related work. Section III presents the PASS design, and Section IV describes the experimental results. Section V summarizes the conclusion. TABLE I concludes the commonly used notations and those descriptions in this paper.

TABLE I     LIST OF ABBREVIATIONS USED IN PAPER

| Notation | Description |
|---|---|
| FL | Federated Learning |
| IoT | Internet of Things |
| FR | Free-Rider |
| AFR | Anonymous Free-Rider |
| SFR | Selfish Free-Rider |
| IID | Independent and Identically Distributed |
| PASS | Parameter Audit-based Secure and fair FL Scheme |
| PASS-CE | Contribution Evaluation in PASS |
| PASS-PPS | Privacy-Preserving Strategy in PASS |
| CE | Contribution Evaluation |
| FedAvg | Federated Averaging |
| SOTA | State-Of-The-Art |
| DSR | Defense Success Rate |
| FPR | False Positive Rate |

## II. RELATED WORK

This section first presents the related work of federated learning and FR attacks. Then, related work on contribution evaluation in FL is presented. At last, related work on defense mechanisms against FR attacks is given.

### A. Free-rider Attack

In FL scenarios, FR adversaries represent a portion of individuals who benefit from a well-trained high-quality global model without contributing computation resources and private data [5], [12], [20]–[22]. The authors in [12] defined AFR adversaries with no privacy dataset and no training ability of a large model. Several studies focused on the AFR attack, such as disguised free-riding [5], novel free-rider [12], and advanced delta weights [22]. In contrast, SFR adversaries are unwilling to devote their privacy dataset and resources to the global model, such as the advanced free-rider attack [12]. Hence, we adopt AFR and SFR attacks to demonstrate the defense capability of PASS.

### B. Contribution Evaluation Methods

Contribution Evaluation (CE) of participants is one of the crucial issues in FL systems because the fair incentive mechanism will evoke the training passion of distributed clients [8]. In addition, unbiased CE will attract clients to devote high-quality and private datasets to FL training [23].

Various researches have been devoted to designing reasonable CE methods [24]–[27]. Zhan et al. [24] summarized two approaches to evaluate the user's contribution to designing incentive mechanisms, including data quantity and quality. Bao et al. [28] proposed FLChain with the third-party auditable method to evaluate the contribution of clients. However, FLChain uses a trusted third party to audit gradient for CE. This method has two weaknesses: (a) if the third party is a free rider, it takes the opportunity to obtain gradient updates and disclose privacy; (b) if the third party is trusted and honest, it is uncertain that the audit result is fair because the performance of a model trained by different datasets is distinct. Liu et al. [26] summarized four CE approaches: (a) self-report the data quality, quantity, and committed computational and communication resources to the server; (b) utility game, which focuses on the changes when a client joins in the FL system; (c) Shapley value, which evaluates the contribution of clients via ablation experiments; (d) individual performance.

The existing CE methods have two weaknesses: (a) They have high requirements for honest and truthful clients, such as self-report and Shapley value solutions. (b) They strongly rely on the participating order of each client, such as the utility game solution. This paper emphasizes that individual performance is the most intuitive and reasonable approach for CE, and we utilize PASS-CE to audit parameters. To reduce the influence of the threshold, we implement theoretical analysis and extensive experiments (in Section IV.E) to explore the most appropriate hyper-parameters.

### C. Defense Mechanisms against FR Attacks

There are two types of defense methods to resist FR attacks: contribution-based and robust model aggregation-based methods. RFFL [8], a contribution-based method, utilizes a reputation mechanism to evaluate the contribution of clients. Median, Trimmed Mean [19], and SignSGD [15] are the representative robust model aggregation-based methods. Yin et al. [19] utilized coordinate-wise median and coordinate-wise trimmed mean instead of weighted averaging. SignSGD is a communication-efficient approach proposed in [15], where participants only upload the element-wise signs of the gradients without the magnitudes.

Notably, there is a critical difference between the above defense mechanisms. Robust model aggregation-based methods tolerate the negative effect of adversaries instead of detecting and removing them. But contribution-based methods, such as RFFL, focus on eliminating adversaries to ensure the security of FL. Furthermore, there are several weaknesses in the methods mentioned above. They cannot handle the situation where attackers occupy more than 50% [14], and only focus on AFR attacks. Moreover, the Median and Trimmed Median significantly reduce the accuracy of complex models on non-iid data and cannot protect the confidentiality of training data [29]. The SignSGD model magnifies the local update, leading to substantially degraded accuracy [30]. In RFFL, using similarities between local and global updates may reduce the contribution of high-quality workers [17]. In this paper, similar to RFFL, our method PASS will eliminate the adversaries from the FL system. In addition, PASS uses reasonable parameter audit for CE and achieves better defense performance than others no matter whether AFR and SFR adversaries occupy the majority of clients or not.

TABLE II          NOTATIONS FREQUENTLY USED IN PASS

| Notation | Description |
|---|---|
| $\eta$ | Learning rate |
| $\varepsilon$ | Gaussian noise |
| $\alpha$ | Moving average coefficient |
| $\beta$ | Threshold coefficient |
| $\gamma$ | Pruning rate |
| $r$ | Round $r$ |
| $R$ | Total number of round |
| $i$ | Client $i$ |
| $N$ | Total number of participants |
| $\sigma_n^2$ | The variance of Gaussian noise, where $n$ is the dimension of the parameter |
| $\theta^0$ | Initialized global model parameter |
| $\tilde{\theta}^0$ | Initialized local model parameter |
| $\tilde{\theta}_i^{r+1}$ | Local model update of the client $i$ |
| $\Delta\tilde{\theta}_{i,ldp}^{r+1}$ | Local model update with weak DP with Gaussian mechanism of the client $i$ in round $r+1$ |
| $AccDiv_i^r$ | Accuracy divergence between $\theta^r$ and $\tilde{\theta}_{i,ldp}^r$ |
| $c_i^r$ | Client $i$ contribution in round $r$ |
| $\dfrac{1}{\beta \times N}$ | Threshold value |

## III. PASS DESIGN

This section first introduces PASS in Section III.A and then details its components, PASS-PPS and PASS-CE, in Section III.B and Section III.C, respectively. Finally, time analysis is presented. TABLE II gives the notations frequently mentioned in PASS.

*A. PASS Description*

PASS aims to establish a secure and fair FL system against FR attacks. It has the following design goals:

**Goal 1.** PASS should protect the privacy dataset of each client from the Deep Leakage from Gradient (DLG) attack [10].

**Goal 2.** PASS should maintain the performance of the FL system in terms of high accuracy and low loss when we adopt **PASS-PPS**.

**Goal 3.** PASS should distinguish between the FR adversaries and fair clients based on the contribution of clients with lower false positive rates.

**Goal 4.** In PASS, the time consumption between clients and the central server should be controlled at an acceptable level.

**Fig. 2** depicts the flow diagram of PASS. After obtaining initialized global model parameters $\theta^0$, each FL client trains $\theta^0$ using a private dataset to obtain the local model update $\tilde{\theta}_i^{r+1}$, where $i \in N$. Then, client $i$ utilizes the **PASS-PPS** to receive the secure local model update $\Delta\tilde{\theta}_{i,ldp}^{r+1}$ (in lines 20-21 of **Algorithm 1**). Moreover, client $i$ uploads $\Delta\tilde{\theta}_{i,ldp}^{r+1}$ to the central server. Notice that $\Delta\tilde{\theta}_{i,ldp}^{r+1}$ is not only used to audit parameters by other clients who possess their privacy dataset but also to achieve model aggregation.

As stated in line 12 of **Algorithm 1**, once received $\Delta\tilde{\theta}_{n,ldp}^{r+1}$ from each client, the server will allocate the new global model parameters $\Delta\theta^{r+1}$ and the local update of each client of the previous round $\Delta\tilde{\theta}_{i,ldp}^{r}$ to achieve parameter audit. The detailed algorithm is described in **Algorithm 1**.

After that, on the client-side, each client audits the received local update using their private dataset and calculates the accuracy divergence $AccDiv_i^r = Acc(\theta^r) - Acc(\Delta\tilde{\theta}_{i,ldp}^r + \tilde{\theta}_{i,ldp}^{r-1})$ of each local update with the previous global update, and we call this parameter audit. On the server side, the server gets the $AccDiv_i^r, i \in N$ and calculates the client contribution $c_i$ by the activation function $tanh(\cdot)$ to map $c_i^r$ to $[-1,1]$, which will divide $c_i^r$ into positive and negative. Once $c_i^r$ is lower than the threshold $\frac{1}{\beta \times N}$, client $i$ will be eliminated from FL system.

To achieve **Goal 1**, **Goal 2**, and **Goal 3**, we need to satisfy Eqs. (1)-(3).

$$\max_{\Delta\tilde{\theta}'} \|D - D_{DLG}\|^2 \quad (1)$$

$$s.t., |Acc - Acc'| \leq \varepsilon \quad (2)$$

$$\arg\min_{\beta,\gamma \in PASS} FPR, \arg\max_{\beta,\gamma \in PASS} DSR \quad (3)$$

In Eq. (1), $\Delta\tilde{\theta}'$ denotes the local update with P**ASS-PPS**. Here, $D$ is the raw input data and $D_{DLG}$ is the data reconstructed by the DLG attack. In Eq. (2), $Acc$ denotes the accuracy of conventional FL while $Acc'$ is that of the FL with **PASS-PPS**, and $\varepsilon$ means a narrow range. In Eq. (3), $DSR$ and $FPR$ are the PASS-CE metrics introduced in Section IV. C.

*B. PASS-PPS*

PASS is a parameter audit-based FL scheme to evaluate the client's contribution to the SFR attack defense. However, in the FL system, the server always adopts the Stochastic Gradient Descent (SGD) optimizer with a learning rate $\eta$. Thus, some adversarial clients can collect the updated gradient to reconstruct the private dataset [10]. That is, the updated gradient is likely to be leaked, according to Eq. (4).

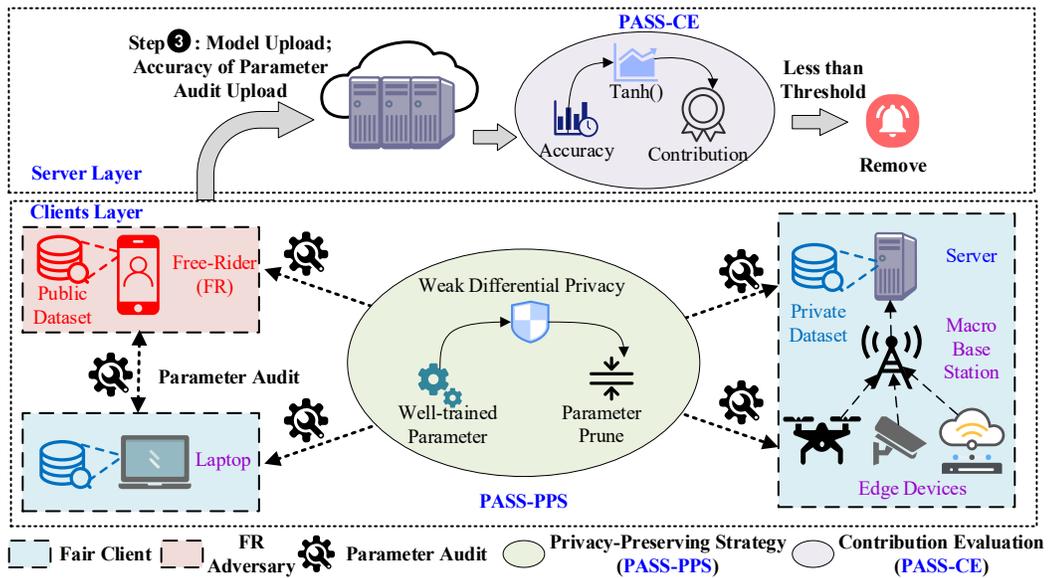

**Fig. 2.** Step (3) in **Fig. 1**: the PASS diagram after being allocated initialized model.

$$\nabla \theta^{r+1} = \frac{\theta^r - \theta^{r+1}}{\eta} \tag{4}$$

With this in mind, and to satisfy **Goal 1**, we utilize **PASS-PPS** to ensure dataset privacy. As described in **Algorithm 1**, after calculating local update $\Delta \tilde{\theta}_i^{r+1}$ by training with the private dataset, the client will utilize weak DP with Gaussian mechanism to add perturbation $\Delta \tilde{\theta}_{i,ldp}^{r+1} = \Delta \tilde{\theta}_i^{r+1} + \varepsilon$, where $\varepsilon \sim \mathcal{N}(0, \sigma_n^2)$. Moreover, $\Delta \tilde{\theta}_{i,ldp}^{r+1}$ will be randomly pruned with the pruning rate $\gamma$ to reduce the communication overhead and guarantee more privacy. Section IV.D demonstrates the achievement of **Goal 1** and **Goal 2**.

*C. PASS-CE*

In CE methods, some researchers adopted cosine similarity to measure the angular distance between updates and to determine the quality of model updates [8], [31], [32]. However, the SFR adversaries will not be recognized effectively in the FL system.

Thus, to satisfy **Goal 3**, we utilize **PASS-CE**, namely validating the other client's local update using the own private dataset on the client-side. Notice that, receiving $AccDiv_i^r$, the server will calculate the averaging $AccDiv_i^r$ using $\left(\frac{1}{n-1}\sum_{i}^{n-1} AccDiv_i^r\right)$ as the fundamental contribution value. Then, the contribution of client $i$ in round $r$ is calculated using the formula $c_i^r = \frac{1}{\alpha \times c_i^{r-1} + (1-\alpha) \times \tanh\left(\frac{1}{N-1}\sum_{i}^{N-1} AccDiv_i^r\right)}$. After that, the server will remove the $i$ where $c_i^r$ lower than the threshold $\frac{1}{\beta \times N}$ (in line 8 of Algorithm 1). Section IV.E indicates the achievement of **Goal 3**.

***Theorem 1 (threshold coefficient)***: Under Algorithm 1, if PASS works, we suppose that the fair client contribution $c_i^r$ satisfies $c_i^r \geq \frac{1}{\beta \times N}$. Namely, given the $\beta \geq \frac{1}{c_i^r \times N}$, we have the threshold coefficient $\beta \geq 1$.

*Proof*: See Appendix.

*D. Time Complexity Analysis*

This section presents time complexity analysis in order to highlight that the pruning strategy can reduce the extra time overhead caused by our PASS.

Frequent communications between server and client will consume a lot of time. Thus, we adopt two strategies to reduce time consumption: (**strategy a**) synchronous transmission; (**strategy b**) parameter pruning.

In (**strategy a**), as stated in line 12 of Algorithm 1, we allocate the global update $\Delta \tilde{\theta}^{r+1}$ together with the local update $\Delta \tilde{\theta}_{\text{client }i,ldp}^r$. And assuming that in a conventional FL system, the communication consumption between the server and each client in each round is $\mathcal{O}(1)$, the time consumption for all clients in all rounds is $\mathcal{O}((N-1) \times N \times R)$ while conventional FL is $\mathcal{O}(1 \times N \times R)$. In the parameter audit phase, the validation consumption is very low in machine learning. Hence, we can ignore the time in each round.

---

**Algorithm 1 The PASS Algorithm**

**Input:** Round $R$, Initial global model parameters $\theta^0$, Client number $N$, Learning rate $\eta$, Threshold coefficient $\beta$, Gaussian noise $\varepsilon$, Pruning rate $\gamma$, Moving average coefficient $\alpha$
**Output:** Update Model Parameters $\theta$
**Initialize:** $\tilde{\theta}^0 = \theta^0$
1:    **For** $r$ in range ($R$):
2:        #========== **Server**==========
3:        **If** $n \in N$: // client $n$ is not eliminated
4:            Allocate $\tilde{\theta}^0$ to client $n$
5:            To step 16
6:        **For** $i$ in range ($N$): // calculate contribution
7:            $c_i^r = \frac{1}{\alpha \times c_i^{r-1} + (1-\alpha) \times \tanh\left(\frac{1}{N-1}\sum_i^{N-1} AccDiv_i^r\right)}$
8:            **If** $c_i^r < \frac{1}{\beta \times N}$: Eliminate $i$ from $N$
9:            **End if**
10:       **End for**
11:       $\Delta \theta^{r+1} = FedAvg(\sum_{i \in N} \Delta \tilde{\theta}_{i,ldp}^{r+1})$ // model aggregation
12:       $(\Delta \theta^{r+1}, \Delta \tilde{\theta}_{\text{client 1},ldp}^r, \Delta \tilde{\theta}_{\text{client 2},ldp}^r, ..., \Delta \tilde{\theta}_{\text{client N},ldp}^r)$ to Client $i$
13:       To step 16 to continue the FL training
14:    **End if**
15:    # ==========**Client** $i$ =========
16:    **If** $r == 0$: Obtain the $\theta^0$ from the Server
17:    **Else**: Obtain the $\Delta \tilde{\theta}^r$ from the Server
18:    **End if**
19:    Calculate local update $\Delta \tilde{\theta}_i^{r+1}$ by training with a private dataset
20:    $\Delta \tilde{\theta}_{i,ldp}^{r+1} = \Delta \tilde{\theta}_i^{r+1} + \varepsilon,\ \varepsilon \sim \mathcal{N}(0, \sigma_n^2)$ // weak DP with Gaussian mechanism
21:    $\Delta \tilde{\theta}_{i,ldp}^{r+1} = (1-\gamma) \cdot \Delta \tilde{\theta}_{i,ldp}^{r+1},\ \gamma \in [0,1)$ // parameter prune
22:    $AccDiv_i^r = Acc(\theta^r) - Acc(\Delta \tilde{\theta}_{i,ldp}^r + \tilde{\theta}_{i,ldp}^{r-1}),\ i \in [1, N-1]$ // Accuracy divergence of previous round using own dataset
23:    Upload $(\Delta \tilde{\theta}_{i,ldp}^{r+1}, AccDiv_1^r, AccDiv_2^r, ..., AccDiv_{N-1}^r)$ to Server // upload the result of parameter audit
24:    To step 6 to continue the FL training
25:  **End for**
26:  **Return** Model Parameters $\theta$

---

In (**strategy b**), motivated by [33], the parameter prune is a practical approach to promote the efficiency of FL training and reduce the communication and computation overhead in FL. Thus, we not only adopt parameter prune as part of **PASS-PPS** but as an effective training accelerating method.

In this situation, the time complexity is $(1-\gamma) \times (N-1)\mathcal{O}(1 \times N \times R)$, notably $\gamma \in [0,1)$. Compared with $\mathcal{O}(1 \times N \times R)$, the communication overhead of PASS is acceptable. Hence, **Goal 4** is satisfied.

## IV. EXPERIMENT EVALUATION

This section first describes the used dataset, baselines, and experiment settings. Then, we introduce the experimental metrics for evaluating FR attack and **PASS**. At last, we demonstrate the experimental results of **PASS-PPS**, which satisfies **Goal 1** and **Goal 2**. We also conduct comparisons with other defense models against AFR and SFR attacks, which satisfies **Goal 3**.

### A. Datasets

MNIST [34] and Cifar10 [35] are used as standard classification baseline datasets. MNIST dataset includes 28x28 handwritten digits with ten classes and has become the most well-known dataset in the classification task. Cifar10 is made up of 10 classes of 32x32 images with three RGB channels and consists of 50,000 training samples and 10,000 testing samples.

Great experiments of studies in FL utilize MNIST and Cifar10 as baseline datasets, so we adopt them in our experiments. Notably, MNIST size is 28x28 while Cifar10 is 32x32, so we assume that FR owns MNIST and honest clients have Cifar10 in FL training. In addition, for the smooth training of the model, we need to extend the same tensor dimension of MNIST as Cifar10.

### B. Baseline

We implement RFFL [8], Median [19], Trimmed Median [19], and SignSGD [15] for comparison in order to reveal PASS's better performance. The experimental settings of each defense model are appropriately utilized.

As stated in [36], the SOTA robust model aggregation rules include Median, Trimmed Median, and SignSGD, which are compared frequently in [8], [32]. On the other hand, RFFL is a SOTA model against AFR attacks. So we implement comparisons in our experiments. Notably, the robust model aggregation-based methods only focus on the overall performance, so we utilize the RFFL threshold standard to remove FR adversaries for evaluating the defense performance in Median, Trimmed Median, and SignSGD. Notably, in [8], RFFL focuses on individual accuracy instead of the fairness of the FL system, so RFFL may not achieve a better performance than others.

### C. Experimental Setting

This subsection gives detailed experimental settings. Then, we introduce the evaluation metrics used in each scenario.

#### 1) Federated Learning

**Models.** We implement a 2-layer convolutional neural network (CNN) [37] for MNIST and a 3-layer CNN [38] for the Cifar10 dataset as the base model in FL.

**Experimental Setting.** The FL is trained via SGD optimizer with learning rate $\eta = 0.1$ and round $r = 200$. FedAvg is adopted as the model aggregation algorithm. In addition, we consider two common types of data, iid and non-iid, in the MNIST and Cifar10 datasets.

#### 2) Free-Rider Attack

**Experimental Setting.** TABLE III gives the number of fair and AFR and SFR adversaries, the AFR and SFR adversary ratio in all clients, the type of data, and the number of samples in training and testing. Notably, we consider the AFR and SFR adversary ratio in three scenarios, including 9%, 33%, 60%, and 67%.

In TABLE IV, the first column denotes the target dataset, namely the data trained by fair clients in FL. The second column is the pre-trained dataset trained by SFR adversaries, representing the SFR adversary unwilling to contribute the privacy dataset to FL model training. Besides, the pre-train dataset MNIST means we extend the same MNIST model tensor dimension as Cifar10 to simulate the SFR attack. In addition, we implement Adam optimizers to conduct the subsequent training in the SFR attack. The last column is the learning rate and decay of Adam.

TABLE III   THE AFR AND SFR ATTACK DETAILS

| Fair Client Num | Adversary Num | Adversary Ratio | Data Split | Train Num | Test Num |
|---|---|---|---|---|---|
| 10 | 1 | 9% | MINST iid/ non-iid | 540 | 60 |
|  | 5 | 33% |  |  |  |
|  | 15 | 60% |  |  |  |
|  | 20 | 67% |  |  |  |
|  | 1 | 9% | Cifar10 iid/ non-iid | 1600 | 400 |
|  | 5 | 33% |  |  |  |
|  | 15 | 60% |  |  |  |
|  | 20 | 67% |  |  |  |

TABLE IV   THE HYPER-PARAMETERS IN SFR ATTACK

| Target Dataset Trained by Fair Client | Pre-trained Dataset Trained by SFR Adversary | | Optimizer | Learning Rate |
|---|---|---|---|---|
| Cifar 10 | iid | MNIST | iid | Adam | 0.015 (Decay: 0.997) |
|  | non-iid |  | non-iid |  |  |

Note: 1) Decay = Learning Rate Decay, which means slowly reducing or decaying the learning rate after each round. 2) Optimizer utilizes the default settings.

#### 3) PASS

**DLG Comparison.** The DLG experimental settings are analogous to that in [10], [16]. We implement L-BFGS [39] optimizer and conduct 300 iterations of optimization to reconstruct the raw data.

**Hyper-Parameters.** In PASS-PPS, we adopt the Gaussian noise distributions $\mathcal{N} \sim (0, \sigma_n^2)$ where the variance satisfies $\sigma_n^2 \in \{0, 10^{-1}, 10^{-2}, 10^{-3}, 10^{-4}, 10^{-5}\}$. Parameter pruning rate $\gamma$ satisfies $\gamma \in [0,1)$. In PASS, the moving average coefficient $\alpha$ is similar to that of [8], namely $\alpha = 0.95$. The threshold coefficient $\beta$ satisfies $\beta \geq 1$, which is proved in APPENDIX. The threshold value of PASS is $\frac{1}{\beta \times N}$, in which $N$ denotes the number of participants in the FL system.

#### 4) Evaluation Metrics

This subsection presents evaluation metrics used in DLG and PASS.

**DLG Evaluation.** We utilize Mean-Square-Error (MSE) between the reconstructed data and raw input to quantify the effectiveness of defenses. A lower MSE indicates a more possibility of data leakage. In addition, we adopt accuracy to reveal the performance loss after using PASS-PPS. It is not acceptable that the accuracy of the whole model decreases sharply. In general, we need to achieve the trade-off with higher MSE and lower accuracy reduction.

**PASS Evaluation.** We define Defense Success Rate (DSR) to reveal the effectiveness of the defense system. DSR, as stated in Eq.(5), denotes the eliminating ratio of FR adversaries in a defense system. Moreover, we adopt a False Positive Rate (FPR), as stated in Eq.(6), to evaluate the performance of attacking defense baselines using AFR and SFR attacks. FPR denotes the removal ratio of fair clients in the detection. F1-score (shown in Eq.(7)) denotes the overall performance between DSR and FPR. In general, a better defense model means lower FPR, higher DSR, and higher F1-score.

$$DSR = \frac{\#\text{Number of Eliminating FR Clients}}{\#\text{Number of All FR Clients}} \times 100\% \quad (5)$$

$$FP = \frac{\#\text{Number of Eliminating Fair Clients}}{\#\text{Number of All Fair Clients}} \times 100\% \quad (6)$$

$$F1-score = \frac{2 \cdot \frac{TP}{TP+FP} \cdot \frac{TP}{TP+FN}}{\frac{TP}{TP+FP} + \frac{TP}{TP+FN}} \times 100\% \quad (7)$$

### D. Evaluation Results of PASS-PPS

This subsection demonstrates the effectiveness of PASS-PPS against the DLG. Motivated by [10], which has proposed the Gaussian noise with variance range from $10^{-1}$ to $10^{-4}$ to defend DLG, we implement weak DP with Gaussian mechanism in experiments. To satisfy **Goal 1** and **Goal 2**, we divide this subsection into two parts. Firstly, to achieve **Goal 1**, we demonstrate the accuracy and loss curves using Gaussian perturbation with various perturbation levels and pruning rates. Then, to satisfy **Goal 2**, we uncover the performance against DLG when adopting the PASS-PPS.

1) **Gaussian weak DP Noise Level and Parameter Prune**

At first, we conduct six noise levels in Gaussian weak DP where variance $\sigma_n^2 \in \{0, 10^{-1}, 10^{-2}, 10^{-3}, 10^{-4}, 10^{-5}\}$. As illustrated in **Fig. 3**, the accuracy gradually decreases with more noise. However, a sharp drop appears when $\sigma_n^2 = 10^{-1}$. Hence, when we add $\varepsilon \sim \mathcal{N}(0, \sigma_n^2)$ where $\sigma_n^2 = 10^{-2}$, the accuracy in MNIST and Cifar10 is similar to that of $\sigma_n^2 = 0$ while the trained data has the maximum privacy. According to **Fig. 4**, the loss curves show a similar result. When $\sigma_n^2 = 10^{-1}$, the loss curve of each data split increases.

Moreover, we implement parameter prune to reduce the leakage of privacy. The pruning rate $\gamma$ satisfies $\gamma \in [0,1)$ and varies from 0 to 90% in experiments. According to **Fig. 5** and **Fig. 6**, we find that the accuracy and loss curves are stable in varying $\gamma$. Thus, to satisfy **Goal 1**, we should select $\sigma_n^2 = 10^{-2}$ no matter which pruning rate is.

2) **PASS-PPS Performance against DLG**

To achieve **Goal 2**, we need to test the PASS-PPS performance against DLG in this subsection. Firstly, we measure the MSE in various noise levels and parameter pruning rates. In **Fig. 7**, the MSE is more than 2.0 when $\sigma_n^2 \geq 10^{-2}$ while MSE is lower than 1.0 when $\sigma_n^2 \leq 10^{-4}$. As evaluated in [16], when MSE is more than 1.49, the system can defend the DLG. **Fig. 8** depicts the MSE comparison with Soteria [16]. If we want to achieve the same defense level as [16], the MSE of PASS-PPS is more than 1.49, which signifies we need select $\sigma_n^2 \geq 10^{-2}$.

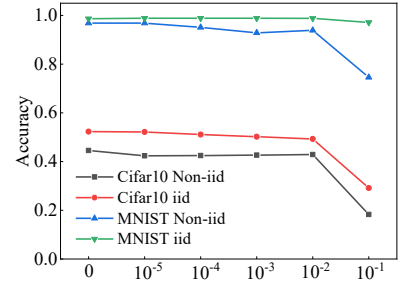

**Fig. 3.** The accuracy with various noise levels $\sigma_n^2$.

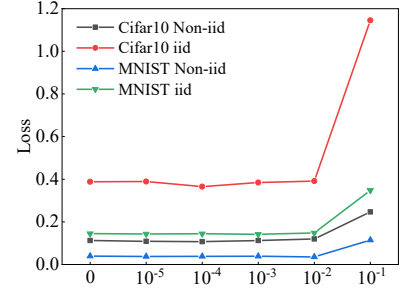

**Fig. 4.** The loss with various noise levels $\sigma_n^2$.

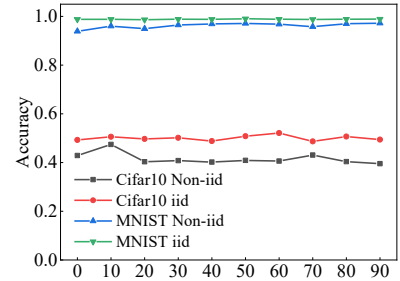

**Fig. 5.** The accuracy with $\gamma$ varying from 0 to 90 when $\sigma_n^2 = 10^{-2}$.

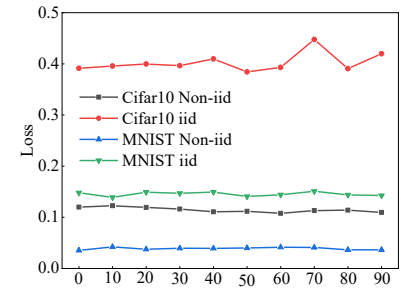

**Fig. 6.** The loss with $\gamma$ varying from 0 to 90 when $\sigma_n^2 = 10^{-2}$.

In conclusion, considering the Gaussian noise level in 1) and the PASS-PPS performance against DLG in 2), we adopt the $\sigma_n^2 = 10^{-2}$ as the Gaussian noise level.

$\gamma = 90\%$ as the pruning rate. In this situation, we achieve 100% DSR and 20% FPR.

TABLE V PASS Performance of DSR and FPR in Varying $\beta$ against SFR attacks

| $\beta$ | Metrics (%) | Number of FR adversary | | | | Avg (%) |
|---|---|---|---|---|---|---|
| | | 1 (9%) | 5 (33%) | 15 (60%) | 20 (67%) | |
| 1.00 | DSR | 100 | 100 | 100 | 100 | 100 |
| | FPR | 65 | 65 | 71 | 100 | 75 |
| | F1-score | 52 | 52 | 45 | 0 | 37 |
| 1.25 | DSR | 100 | 100 | 100 | 100 | 100 |
| | FPR | 38 | 43 | 43 | 100 | 56 |
| | F1-score | 77 | 73 | 73 | 0 | 56 |
| 1.50 | DSR | 100 | 100 | 60 | 100 | 90 |
| | FPR | 32 | 37 | 27 | 40 | 34 |
| | F1-score | 81 | 77 | 66 | 75 | 75 |
| 1.75 | DSR | 100 | 100 | 40 | 100 | **85** |
| | FPR | 23 | 20 | 19 | 30 | **23** |
| | F1-score | 87 | 89 | 60 | 82 | **80** |
| 2.00 | DSR | 100 | 100 | 20 | 0 | 55 |
| | FPR | 20 | 20 | 13 | 10 | 16 |
| | F1-score | 89 | 89 | 57 | 46 | 70 |

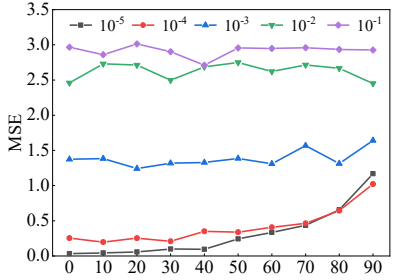

**Fig. 7.** The MSE of DLG in varying $\gamma$ and $\sigma_n^2$ in Cifar10.

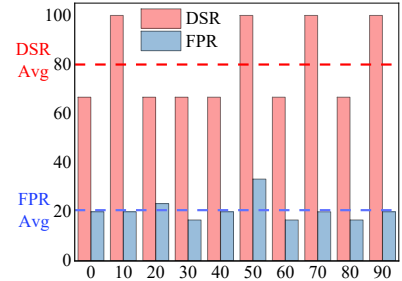

**Fig. 8.** The MSE comparison with Soteria [16] against DLG.

### E. Selection of $\beta$ and $\gamma$ in PASS-CE

In APPENDIX, we prove that the threshold coefficient $\beta$ needs to satisfy $\beta \geq 1$. In this subsection, we investigate the PASS performance of varying $\beta$ and $\gamma$.

To find the best $\beta$, we first implement varying $\beta$ under the pruning rate $\gamma \in (0,1]$. TABLE V shows the experiment results with ten fair clients and 1, 5, 15, and 20 SFR adversaries under $\gamma \in (0,1]$ and varying $\beta$. We observe that when $\beta = 1.00$, PASS can 100% defend SFR attack, but the FPR is 75% which means 75% of fair clients have been eliminated from the FL system by mistake. On the other hand, with $\beta = 2.00$, PASS achieves 16% FPR, but the DSR is 55% meaning only 55% of SFR adversaries are eliminated from the FL system. Apparently, those situations are not proper in the defense model. At last, we select $\beta = 1.75$ as the threshold coefficient value. In this situation, PASS achieves a DSR of 85%, an FPR of 23%, and an F1-score of 80%, reaching the trade-off between DSR and FPR.

To investigate the effect of varying $\gamma$, we fix $\beta = 1.75$ and conduct extensive experiments. In **Fig. 9**, the red dashed line denotes the average DSR of 80%, and the blue dashed line indicates the average FPR of 21%. The red bar represents the DSR, and the blue bar represents the FPR with varying $\gamma$. To satisfy **Goal 3**, we must achieve a trade-off between DSR and FPR. With this in mind, we choose $\gamma = 10\%$, $\gamma = 70\%$, and $\gamma = 90\%$. However, to satisfy **Goal 4**, we need to utilize larger $\gamma$ to reduce communication consumption. Hence, we adopt

**Fig. 9.** DSR and FPR in varying $\gamma$ when $\beta = 1.75$.

To demonstrate the effectiveness of PASS, we provide three groups of experiments with $\beta = 1.75$ and $\gamma = 90\%$, which include no defense scenario and PASS scenario with 1, 5, 15, and 20 SFR adversaries. As illustrated in **Fig. 10**, the red dashed line is the PASS worked round. Compared with the no-defense scenario, the PASS scenario eliminated the SFR adversaries and increased accuracy. Notably, PASS curves differ because of the existing FPR of 10% in 1 SFR adversary, 30% in 5 SFR adversaries, 20% in 15 SFR adversaries, and 20% in 20 SFR adversaries.

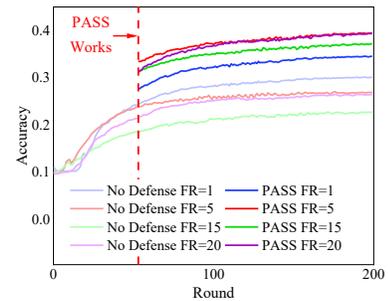

**Fig. 10.** The accuracy curves of PASS and no defense.

### F. PASS Performance Evaluation

In the previous section, we confirm all hyper-parameters. As a result, this subsection investigates the PASS performance compared with other defense models, including RFFL, Median, Trimmed Median, and SignSGD.

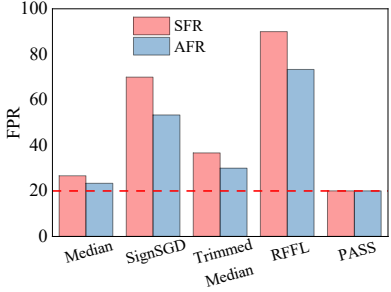

**Fig. 11.** The FPR of defense models against AFR and SFR attacks.

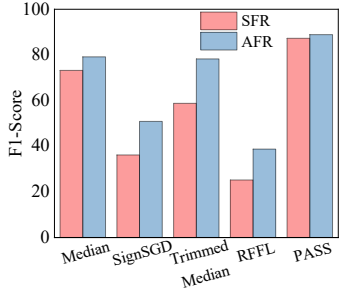

**Fig. 12.** The F1-score of defense models against AFR and SFR attacks.

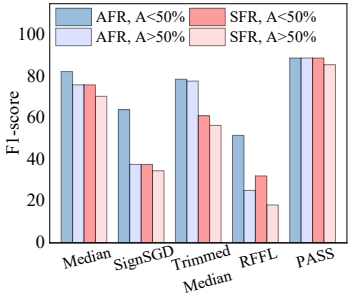

**Fig. 13.** The F1-score of defense models against AFR and SFR attacks with different adversary proportions.

We adopt proper experimental settings for each defense model demonstrated in Section IV. C. **Fig. 11**, **Fig.12,** and **Fig.13** illustrate the three-fold observations: (1) against the AFR attack and the SFR attack, PASS we proposed achieves the lowest FPR, and the highest F1-score; (2) the SFR attack scenario has a better attack performance than the AFR attack; (3) compared with other schemes, PASS has the best F1-score whether adversary number (A in **Fig.13**) is more than 50% or less than 50%.

In fact, each defense model can reach almost 100% DSR by using proper hyper-parameters in AFR and SFR attacks, but they can not reach a better trade-off between DSR and FPR. However, our PASS achieves the lowest FPR with 100% DSR and obtains the highest F1-score (on average) of 88% against SFR and 89% against AFR.

## V. CONCLUSION

Free-Rider (FR) attacks enable the FR adversary to benefit from the well-trained model without contributing any private dataset and computation resources in Federated Learning (FL). Against AFR and SFR attacks, this paper proposes a *P*arameter *A*udit-based *S*ecure and fair FL *S*cheme (PASS) to evaluate participant contribution intuitively. To protect private data during parameter auditing, we adopt a privacy-preserving strategy (PASS-PPS) that utilizes weak Differential Privacy with a Gaussian mechanism and parameter prune mechanism. For achieving fair FL training, we utilize the new contribution evaluation method to measure individual performance.

Extensive experiments are conducted to evaluate the performance of PASS. Our PASS-PPS shows a similar defense level as a state-of-the-art model against privacy leakage. In AFR and SFR attacks, PASS has a 100% Defense Success Rate (DSR), the lowest False Positive Rate (FPR), and the highest F1-score compared with other defense models. Besides, the experiment results demonstrate that PASS produces no negative effect on FL accuracy when there is no FR adversary.

Note that in this paper we use a standalone framework to simulate FL on a hardware device, which can not measure communication cost. Therefore, there is no experiment for evaluating the effectiveness of synchronous transmission and parameter prune in reducing communication overhead caused by frequent communications between the server and clients in the parameter audit phase. In our future work, we plan to establish distributed training experiments in order to evaluate communication costs.

## VI. ACKNOWLEDGEMENT

The research of Jianhua Wang, Xiaolin Chang, and Yixiang Wang was supported in part by National Natural Science Foundation of China under Grant No.62272028. The work of Jelena Mišić and Vojislav B. Mišić was supported by Natural Science and Engineering Research Council (NSERC) of Canada. The research of Jianhua Wang was also supported by the Fundamental Research Funds for the Central Universities 2022YJS030.

## APPENDIX

### PROOF OF THEOREM 1 IN SECTION III. C

According to **Algorithm 1**, we evaluate the contribution of the client $i$ is $c_i^r = \dfrac{1}{\alpha \times c_i^{r-1} + (1-\alpha) \times \tanh\left(\dfrac{1}{N-1}\sum_{i}^{N-1} AccDiv_i^r\right)}$, and the threshold value is $\dfrac{1}{\beta \times N}$. If client $i$ can be the honest participant, the contribution should satisfy $c_i^r \geq \dfrac{1}{\beta \times N}$. As a result, we obtain the relationship between $c_i^r$ and threshold $\dfrac{1}{\beta \times N}$. In addition, assuming that $c_i^r$ is similar with $c_i^{r-1}$ within one single round, we get the equation $c_i^r \times \left[\alpha \times c_i^r + (1-\alpha) \times \tanh\left(\dfrac{1}{n-1}\sum_{i}^{n-1} AccDiv_i^r\right)\right] = 1$.

In this situation, we solve that equation with the following roots: $c_i^r = \dfrac{-(1-\alpha)\tanh\left(\dfrac{1}{N-1}\sum_{i}^{N-1} AccDiv_i^r\right) \pm z}{2\alpha}$, where

$z = \sqrt{\left[(1-\alpha)\tanh\left(\dfrac{1}{N-1}\sum_{i}^{N-1} AccDiv_i^r\right)\right]^2 + 4\alpha}$.

Since $c_i^r$ should be positive, we simultaneously get the relationship $c_i^r \geq \dfrac{1}{\beta \times N}$ and receive the relationship: $\beta \geq \dfrac{1}{c_i^r \times N}$, where $\tanh\left(\dfrac{1}{N-1}\sum_{i}^{N-1} AccDiv_i^r\right) \in [-1,1]$, and the number of all clients $N \in [1,25]$ in our scenario. So, we adopt the boundary $\tanh\left(\dfrac{1}{N-1}\sum_{i}^{N-1} AccDiv_i^r\right) = 1$ and $N = 1$. In this situation, we can obtain $\beta \geq \dfrac{2\alpha}{\alpha - 1 + \sqrt{(1-\alpha)^2 + 4\alpha}}$, which can be simplified to be $\beta \geq \dfrac{2\alpha}{\alpha - 1 + 1 + \alpha}$. Thus, we prove the infimum of the threshold coefficient $\beta \geq 1$. ∎

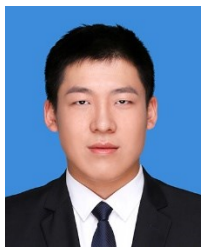

Jianhua Wang received the B.S. and M.S. degrees in Software engineering from the Taiyuan University of Technology in 2017 and 2020. He now pursues his Ph.D. degree at Beijing Jiaotong University, majoring in Cyberspace Security. His research interests include adversarial machine learning and federated learning.

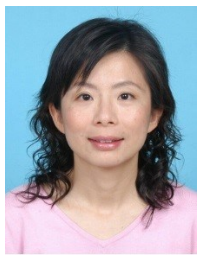

Xiaolin Chang is Professor at School of Computer and Information Technology, Beijing Jiaotong University. Her current research interests include Cloud-edge computing, cybersecurity, secure and dependable in machine learning. She is a member of IEEE.

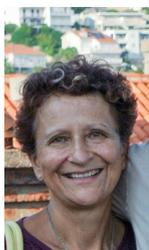

Jelena Mišić is Professor of Computer Science at Ryerson University in Toronto, Ontario, Canada. She has published papers in archival journals and at international conferences in the areas of wireless networks, in particular wireless personal area network and wireless sensor network protocols, performance evaluation, and security. She serves on editorial boards of IEEE Transactions on Vehicular Technology, Computer Networks, Ad hoc Networks, Security and Communication Networks, Ad Hoc & Sensor Wireless Networks, Int. Journal of Sensor Networks, and Int. Journal of Telemedicine and Applications. She is a Fellow of IEEE and Member of ACM.

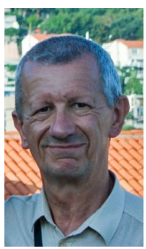

Vojislav B. Mišić is Professor of Computer Science at Ryerson University in Toronto, Ontario, Canada. He received his PhD in Computer Science from University of Belgrade, Serbia, in 1993. His research interests include performance evaluation of wireless networks and systems and software engineering. He has authored or co-authored six books, 20 book chapters, and over 280 papers in archival journals and at prestigious international conferences. He serves on the editorial boards of IEEE transactions on Cloud Computing, Ad hoc Networks, Peer-to-Peer Networks and Applications, and International Journal of Parallel, Emergent and Distributed Systems. He is a Senior Member of IEEE and member of ACM.

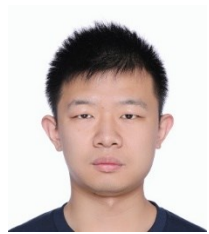

Yixiang Wang received his B.S. (2018) degree from Beijing Jiaotong University, China. He is a Ph.D. candidate at Beijing Key Laboratory of Security and Privacy in Intelligent Transportation, Beijing Jiaotong University. His research interests include adversarial examples and security in machine learning.